\documentclass[runningheads]{llncs}
\usepackage{graphicx}
\usepackage{amsfonts}
\usepackage{xcolor}
\usepackage{verbatim}
\usepackage{makecell}
\begin{document}

\title{Self-Supervised Generative Adversarial Network for Depth Estimation in Laparoscopic Images}
%
%
\author{Baoru Huang\inst{1,2} \and
Jianqing Zheng\inst{3} \and
Anh Nguyen\inst{1}  \and
David Tuch \inst{4}  \and
Kunal Vyas \inst{4}  \and
Stamatia Giannarou \inst{1,2}  \and
Daniel S. Elson \inst{1,2}}
%
\institute{The Hamlyn Centre for Robotic Surgery, Imperial College London, SW7 2AZ, UK \\
\email{Baoru.Huang18@imperial.ac.uk} \and
Department of Surgery $\&$ Cancer, Imperial College London, SW7 2AZ, UK \and
The Kennedy Institute of Rheumatology, University of Oxford, UK \and
Lightpoint Medical Ltd.
\\}

\maketitle              

\begin{abstract}
Dense depth estimation and 3D reconstruction of a surgical scene are crucial steps in computer assisted surgery. Recent work has shown that depth estimation from a stereo images pair could be solved with convolutional neural networks. However, most recent depth estimation models were trained on datasets with per-pixel ground truth. Such data is especially rare for laparoscopic imaging, making it hard to apply supervised depth estimation to real surgical applications. To overcome this limitation, we propose SADepth, a new self-supervised depth estimation method based on Generative Adversarial Networks. It consists of an encoder-decoder generator and a discriminator to incorporate geometry constraints during training. Multi-scale outputs from the generator help to solve the local minima caused by the photometric reprojection loss, while the adversarial learning improves the framework generation quality. Extensive experiments on two public datasets show that SADepth outperforms recent state-of-the-art unsupervised methods by a large margin, and reduces the gap between supervised and unsupervised depth estimation in laparoscopic images.

\keywords{Depth Estimation  \and Laparoscopic Images \and Generative Adversarial Network}
\end{abstract}

\section{Introduction}
Robot-assisted minimally invasive surgery with stereo laparoscopic vision has become popular due to the advantages of enhanced movement range, precision, vision and proficiency \cite{mack2001minimally,nguyen2020end,zhang2018self}. Surgical scene depth estimation is a fundamental problem in image-guided intervention and has received substantial prior interest to its promise for robot navigation, 3D registration between pre- and intra-operative organ models, and augmented reality \cite{ye2017self}. Obtaining depth maps is not trivial due to the inherent problems such as tissue deformation, specular reflections, and lack of photometric constancy across frames \cite{liu2019dense}. 

Several traditional methods used multi-view stereo algorithms such as Simultaneous Localization and Mapping (SLAM) \cite{grasa2013visual} and Structure from Motion (SfM) \cite{leonard2018evaluation}, but these struggle with less textured tissues. More recently deep learning-based depth estimation has used RGB images as the training data and Convolutional Neural Networks (CNNs) for supervised learning  \cite{eigen2014depth,do2021multiple}. To produce accurate results in less than a second of GPU time, Luo \textit{et al.} \cite{luo2016efficient} treated the problem as a multi-class classification indicating all possible disparities, and exploited a product layer to simplify the representations of a Siamese architecture. Chang \textit{et al.} \cite{chang2018pyramid} proposed PSMNet, where the capacity of global context information at different scales and locations could be extracted by a spatial pyramid pooling module to form a cost volume. Duggal \textit{et al.} \cite{duggal2019deeppruner} sped up the runtime of stereo matching and developed a differentiable PatchMatch module that could discard most disparities without the need of full cost volume evaluation. 

The methods above are fully supervised and require ground truth depth during training. However, acquiring per-pixel ground truth depth data is challenging for real-world settings \cite{joung2019unsupervised} and especially for laparosocpic vision where port space is limited, working distance is short and sterilization is required \cite{huang2020tracking}. One alternative is self-supervised training of depth estimation models using image reconstruction as the supervisory signal \cite{garg2016unsupervised}. The input is usually a set of images in the form of monocular or stereo images \cite{zhou2017unsupervised}. Godard \textit{et al.} \cite{godard2017unsupervised} proposed a training loss that included a left-right depth consistency term and a reconstruction term for single image depth estimation, despite the absence of ground truth depth. This was extended by \cite{godard2019digging} with full-resolution multi-scale sampling to reduce visual artifacts, and a minimum reprojection loss to robustly handle occlusions. Johnston \textit{et al.} \cite{johnston2020self} further closed the gap with fully-supervised methods by including a self-attention mechanism and made use of contextual information. Ye \textit{et al.} \cite{ye2017self} proposed a deep learning framework for surgical scene depth estimation in self-supervised mode for scalable data acquisition by adopting a differentiable spatial transformer and an autoencoder. 

In this paper, we present a new method for self-supervised adversarial depth estimation: SADepth. A U-Net architecture \cite{ronneberger2015u} was adopted as a generative structure and fed with stereo pairs as inputs to benefit from complementary information. To cope with local minima caused by classic photometric reprojection loss, we applied the disparity smoothness loss and formed the network across multiple scales. The use of a generative adversarial network (GAN) allowed us to improve the reconstructed image quality, which formed a supervisory signal for training, while keeping the overall end-to-end optimization objective.

\section{Methodology}

\subsection{Overview}
Here we describe the proposed self-supervised adversarial depth estimation framework, SADepth. Stereo depth estimation predicts depth maps $\textbf{\textit{D}}^{\rm l},\textbf{\textit{D}}^{\rm r}\in\mathbb{R}_{+}^{h\times w}$ based on the stereo RGB images ${\textbf{\textit{I}}^{\rm l}},{\textbf{\textit{I}}^{\rm r}}\in\mathbb{R}_{+}^{h\times w\times 3}$ of height and width $h,w$. A generative network $\mathcal{G}$ with stereo image pairs ${\textbf{\textit{I}}^{\rm l}}$ and ${\textbf{\textit{I}}^{\rm r}}$ as inputs, was used to produce two distinct left and right disparity maps ${\textbf{\textit{d}}^{\rm l}}$ and ${\textbf{\textit{d}}^{\rm r}}$, \textit{i.e.} ${\textbf{\textit{d}}^{\rm l}}$, ${\textbf{\textit{d}}^{\rm r}}$ = ${\mathcal{G}(\textbf{\textit{I}}^{\rm l}, \textbf{\textit{I}}^{\rm r})}$. As the two disparity maps were generated from different input images, a `reprojection sampler' \cite{jaderberg2015spatial} could be used for photometric reprojection loss computation of mutual counter-parts, \textit{i.e.} reconstructed left and right images ${\textbf{\textit{I}}^{\rm l*}}$ and ${\textbf{\textit{I}}^{\rm r*}}$ . The discriminator $\mathcal{D}$ was exploited to indicate if the reconstructed images were real or fake (original input images were regarded as real). By forcing the reconstructed image to be consistent with the original input, we could derive accurate disparity maps for depth inference, as shown in the following sections.

\begin{figure}[t]
\includegraphics[scale=0.38]{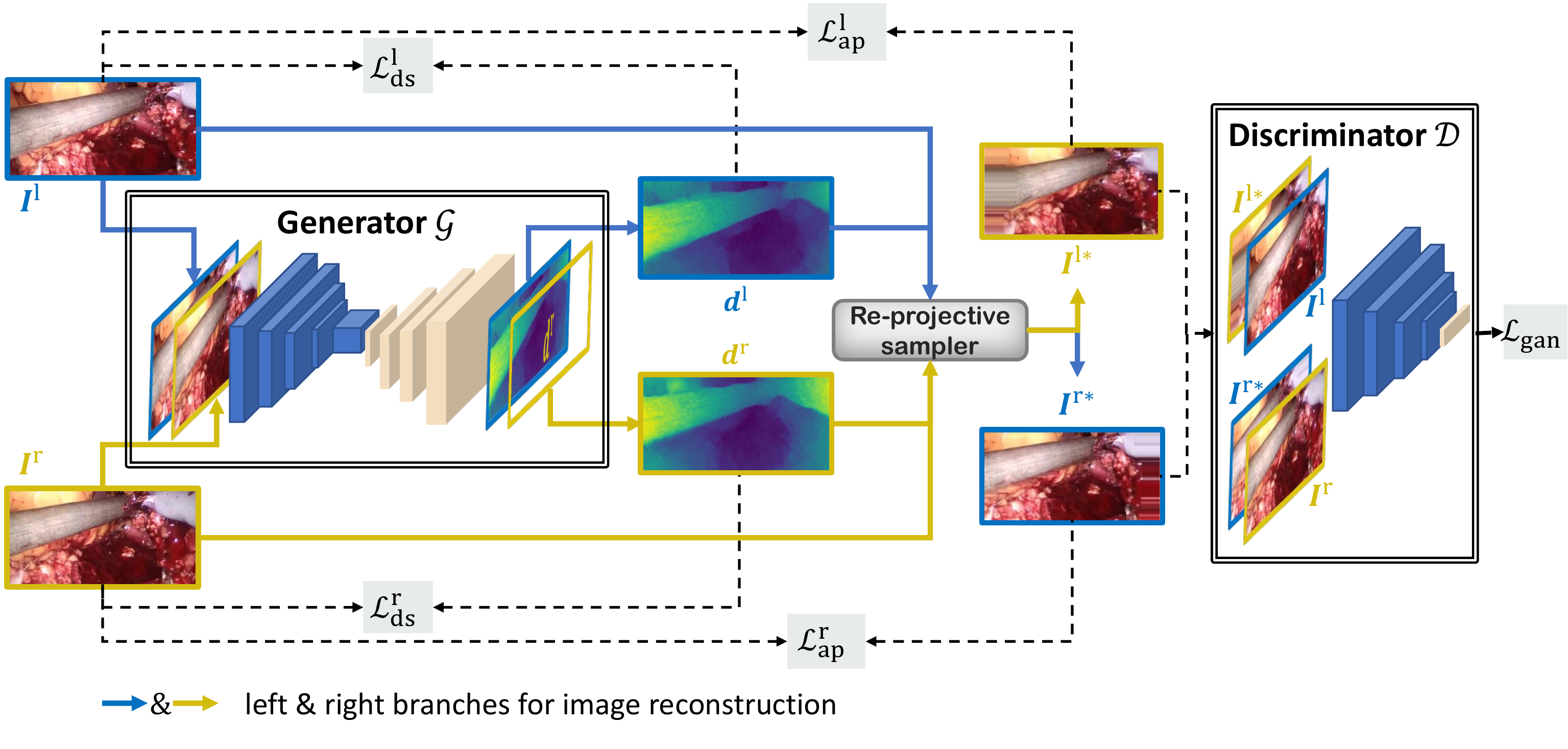}
\caption{Overview of the self-supervised adversarial depth estimation network, SADepth. 
} \label{overview}
\end{figure}

\subsection{Network Architecture}
\paragraph{Generator.}
The generator followed the general U-Net \cite{ronneberger2015u} architecture consisting of an encoder-decoder network, where the encoder was designed to obtain compact image representations and the decoder produced disparity maps for left and right input images, recovering them at the original scale (illustrated in Figure \ref{network}). Encoder-decoder skip connections were applied to represent deep abstract features while preserving local information. To make the model compact - and different from less streamlined previous approaches which had two branches or two sub-networks for the encoder \cite{chang2018pyramid} \cite{pilzer2018unsupervised} \cite{huang2021h} - we first concatenated the left and right images into a 6-channel tensor and then fed it to a ResNet18 model \cite{he2016deep}. The input size was \(\emph{\# channels}\times h\times w=6 \times 192 \times 384\). Similar to \cite{godard2017unsupervised}, our decoder was formed of five cascaded blocks where each block had four parts: the first convolutional layer, an upsampling layer, a concatenation manipulation, and the second convolutional layer. In the upsampling layer, features were interpolated to twice the input size and both convolutional layers were followed by an \textit{ELU} activation function \cite{clevert2015fast}. In particular, sigmoids were applied at the output to generate a 2-channel tensor representing the left and right disparity ${\textbf{d}^{\rm l}}$ and ${\textbf{d}^{\rm r}}$. Finally the sigmoid outputs were converted to depth by \(\textbf{D}^{l(r)} = 1/(a \textbf{d}^{l(r)} + b)\), where parameters \(a\) and \(b\) were selected to constrain the depth \(\textbf{D}^{l(r)}\) between 0.1 and 100 units. The depth maps were then back-projected into point clouds by applying the intrinsic parameters and using the counter-part camera's extrinsic parameters to form reconstructed stereo images. The structural similarity between the original and reconstructed images was regarded as a supervisory signal to train the generator (see section 2.3 for the generator loss).

\paragraph{Discriminator.}
Godfellow \textit{et al.} \cite{goodfellow2014generative} introduced a generative adversarial learning strategy and presented impressive results for image generation tasks. GANs have been widely exploited in different tasks with different GAN models including \textit{e.g.} DualGAN \cite{yi2017dualgan} and CycleGAN \cite{zhu2017unpaired}. To improve the generation quality of the reconstructed images ${\textbf{\textit{I}}^{\rm l*}}$ and ${\textbf{\textit{I}}^{\rm r*}}$, and following the work in \cite{pilzer2018unsupervised} for natural scenes, we applied an adversarial learning strategy for laparoscopic images to include geometry constraints during training and force the network to make a consistent depth map prediction. The original input stereo image pairs and reconstructed images \(\textbf{\textit{I}}^{\rm r*}\) and \(\textbf{\textit{I}}^{\rm l*}\) generated from the `reprojection sampler' were fed into the discriminator \(\mathcal{D}\), which consisted of convolutional, batch normalization and activation function layers and classified the input and reconstructed images as real or fake. As training progressed, the reconstructed images became more similar to the original inputs, while the discriminator also became better at distinguishing between the input and reconstructed images, resulting in an overall improvement of the associated disparity maps. 


\begin{figure*}[t]
\includegraphics[width=\textwidth]{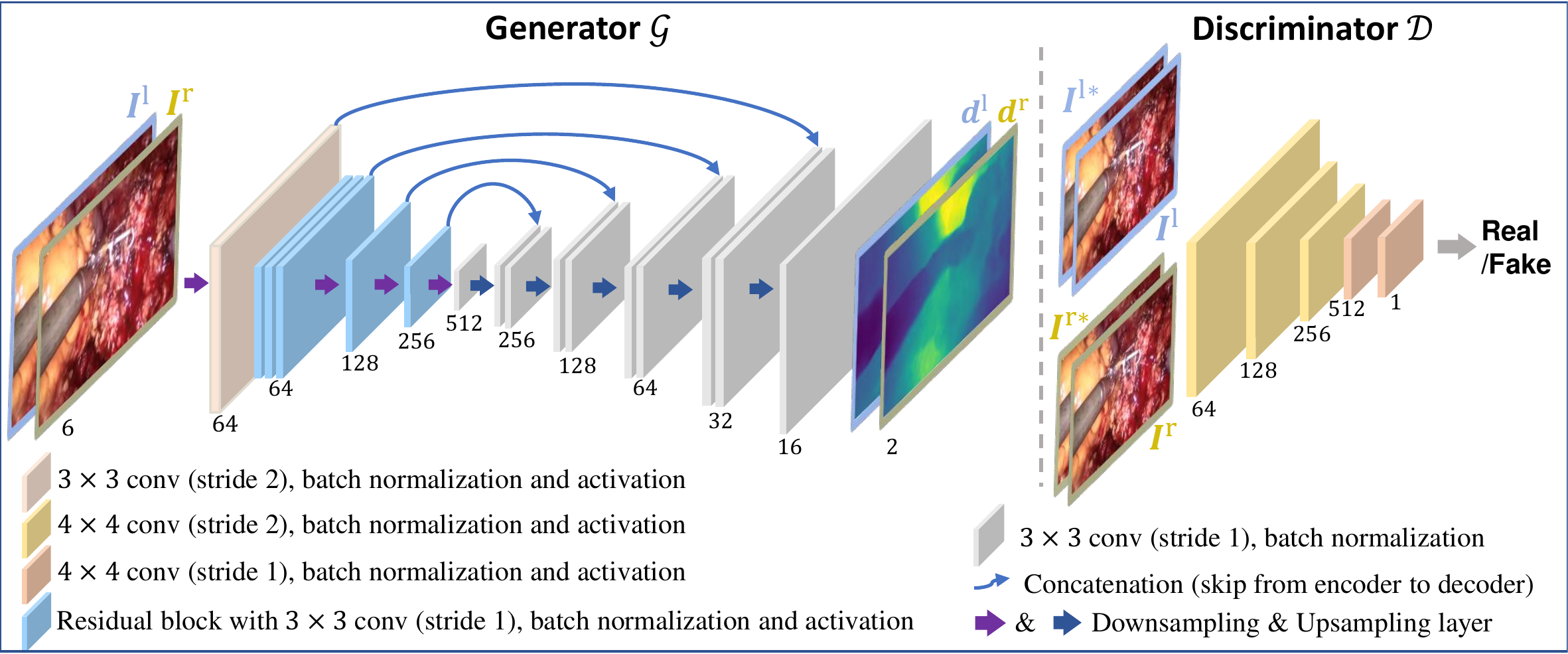}
\caption{The detailed architecture of the SADepth generator and discriminator. The generator was an autoencoder architecture with concatenated stereo image pairs as inputs and left and right disparity maps as outputs using a sigmoid function. These outputs were then transformed to reconstruct the counter-part camera input images using a `reprojection sampler', and these reconstructed images were fed into the discriminator together with the original input image pair. The discriminator output a scalar indicating whether the reconstructed images generated from the `reprojection sampler' were real or fake.} 
\label{network}
\end{figure*}

\subsection{Training Losses}
\paragraph{Generator Loss.}  

In the depth estimation generator network \(\mathcal{G}\), the loss \(\mathcal{L}_{\rm rec}^{\rm r}\) was formed from the appearance matching loss \(\mathcal{L}_{\rm ap}^{\rm r}\) and disparity smoothness loss \(\mathcal{L}_{\rm ds}^{\rm r}\) 
\begin{equation}
\mathcal{L}_{\rm rec}^{\rm r} = \mathcal{L}_{\rm ap}^{\rm r} + \alpha_{\rm ds} \mathcal{L}_{\rm ds}^{\rm r}
\end{equation}
where \(\alpha_{\rm ds}\) balanced the loss magnitude of the two parts to stabilize the training and was set to 0.001.

\paragraph{Appearance-Matching Loss.}
Self-supervised training typically assumes that the appearance and material properties (\textit{e.g.} brightness and Lambertian) of object surfaces are consistent between frames. A local structure-based appearance loss \cite{godard2017unsupervised} can effectively improve the depth estimation performance compared with simple pairwise pixel differences \cite{zhou2017unsupervised}. Following \cite{godard2019digging}, we exploited the appearance-matching loss as part of the generator loss which forced the reconstructed image to be similar to the corresponding training inputs. 
During the training, the right disparity map ${\textbf{d}^{\rm r}}$ generated by the autoencoder was then transformed to produce ${\textbf{\textit{I}}^{\rm r*}}$ -- a reconstruction of the original right input image -- using RGB intensity information from the counter-part camera image ${\textbf{\textit{I}}^{\rm l}}$ (see Fig.~\ref{overview}). This was achieved by first converting the disparity map ${\textbf{d}^{\rm r}}$ to a depth map ${\textbf{D}^{\rm r}}$, from which a point cloud of the surgical scene could be generated. Then the point cloud was transferred into the other camera's coordinate system and projected onto its image plane. The reconstructed input image  ${\textbf{\textit{I}}^{\rm r*}}$ was generated with bilinear interpolation for each output pixel using the weighted sum of the four neighboring intensities. In contrast to \cite{garg2016unsupervised}, this bilinear sampling was locally fully differentiable, which allowed it to be integrated into the fully convolutional architecture without requiring simplification or approximation of the cost function. To compare the reconstructed image ${\textbf{\textit{I}}^{\rm r*}}$ and the original input image ${\textbf{I}^{\rm r}}$, a combination of structural similarity (SSIM) index \cite{wang2004image} and ${\mathcal{L}_1}$ loss were applied as the photometric image reconstruction cost ${\mathcal{L}_{ap}^{\rm r}}$: 
\begin{equation} \label{eq:appearance_matching_loss}
\mathcal{L}_{ap}^{\rm r} = \frac{1}{N}\sum_{i, j} \frac{\gamma}{2} (1-{\rm SSIM}(I_{ij}^{\rm r}, I_{ij}^{r*})) + (1-\gamma) {\|I_{ij}^{\rm r}-I_{ij}^{r*}\|}_1
\end{equation}
where \(N\) denotes the number of pixels and \(\gamma\) represents the weighting for L1-norm loss term, which was set to 0.85. Similar to \cite{godard2017unsupervised}, the calculation of SSIM here was simplified to a \(3\times 3\) block filter instead of a Gaussian. The training of the depth estimation generator then involved minimizing the reconstruction loss between input and reconstructed images.

\paragraph{Disparity Smoothness Loss.}
Since disparities should be locally smooth and discontinuities usually occur at image gradients, we applied the disparity smoothness loss to penalize unexpected discontinuities in the disparity maps. Following \cite{heise2013pm}, this cost was an edge-aware term weighted with the input image gradients \(\partial \textbf{I}\):

\begin{equation}
\mathcal{L}_{\rm ds}^{\rm r} = \frac{1}{N}\sum_{ij}|\partial_x (\textbf{d}_{ij}^{\rm r})|e^{-|\partial_x \textbf{\textit{I}}_{ij}^{\rm r}|} + |\partial_y (\textbf{d}_{ij}^{\rm r})| e^{-|\partial_y \textbf{\textit{I}}_{ij}^{\rm r}|}
\end{equation}
where \(\textbf{d}^{\rm r}\) represents the generated disparity map and \(\textbf{\textit{I}}^{\rm r}\) is the original input right image. 

\paragraph{Discriminator Loss.}  
The adversarial objective of the generative network can be expressed as follows:
\begin{equation}
\label{GAN}
\mathcal{L}_{\rm gan}^{\rm r}(\textbf{\textit{I}}^{\rm r}, \textbf{\textit{I}}^{\rm r*};\mathcal{G}, \mathcal{D}) = \mathbb{E}_{\textbf{\textit{I}}^{\rm r}\sim P(\textbf{\textit{I}}^{\rm r})} [{\rm log} (\mathcal{D}(\textbf{\textit{I}}^{\rm r}))] + \mathbb{E}_{\textbf{\textit{I}}^{\rm r*}\sim P(\textbf{\textit{I}}^{\rm r*})} [{\rm log}(1-\mathcal{D}(\textbf{\textit{I}}^{\rm r*}))]
\end{equation}
where a cross-entropy loss measured the expectation of the reconstructed image \(\textbf{\textit{I}}^{\rm r*}\) against the distribution of the input image \(\textbf{\textit{I}}^{\rm r}\). 
Note that both generator and discriminator losses included losses for left and right images but only the right image equations are shown.

\paragraph{Multi-Scale Loss.}
One remaining issue with the above learning pipeline was that the training objective risked becoming stuck in local minima due to the application of a photometric reprojection loss \cite{watson2019self}.
The strategy introduced in \cite{zhou2017unsupervised} indicated that combining the individual losses across multiple scales in the decoder was effective, which could improve the depth estimation performance and reduce sensitivity to architectural choices. Hence, the lower resolution depth maps (from the intermediate layers) were first upsampled to the input image resolution and then reprojected and resampled, with the errors computed at the higher input resolution. This manipulation is similar to matching patches, which enables low-resolution disparity maps to warp an entire patch of pixels in a high resolution image while promoting the depth maps at every scale to reconstruct the high resolution input image as accurately as possible \cite{godard2019digging}.

\paragraph{Joint Optimization Loss} Finally, the joint optimization loss was a combination of generator loss and adversarial loss, written as:
\begin{equation} \label{eq:total_depth_loss}
\mathcal{L}_{\rm total} =\frac{1}{m}\sum_{s=1}^m \frac{\mathcal{L}_s^{\rm l} + \mathcal{L}_s^{\rm r}}{2}=  \frac{1}{m} \sum_{s=1}^m \big(\alpha(\mathcal{L}_{\rm rec}^{\rm l} + \mathcal{L}_{\rm rec}^{\rm r}) + \beta (\mathcal{L}_{\rm gan}^{\rm l} + \mathcal{L}_{\rm gan}^{\rm r})\big)
\end{equation}

\subsubsection{Training}
The depth estimation procedure was trained based on the reconstruction supervision signal and no per-pixel depth ground truth labels were needed. The augmentation of input data was performed on the fly by flipping 50 \% of the input images horizontally and reorienting the stereo pairs. Parameter $m$ was set to 4, which means that there were 4 output scales with resolutions \(\frac{1}{2^0}\), \(\frac{1}{2^1}\), \(\frac{1}{2^2}\) and \(\frac{1}{2^3}\) of the input resolution. \(\alpha\) and \(\beta\) were set to 0.5.


\begin{figure*}[t]
\includegraphics[width=\textwidth]{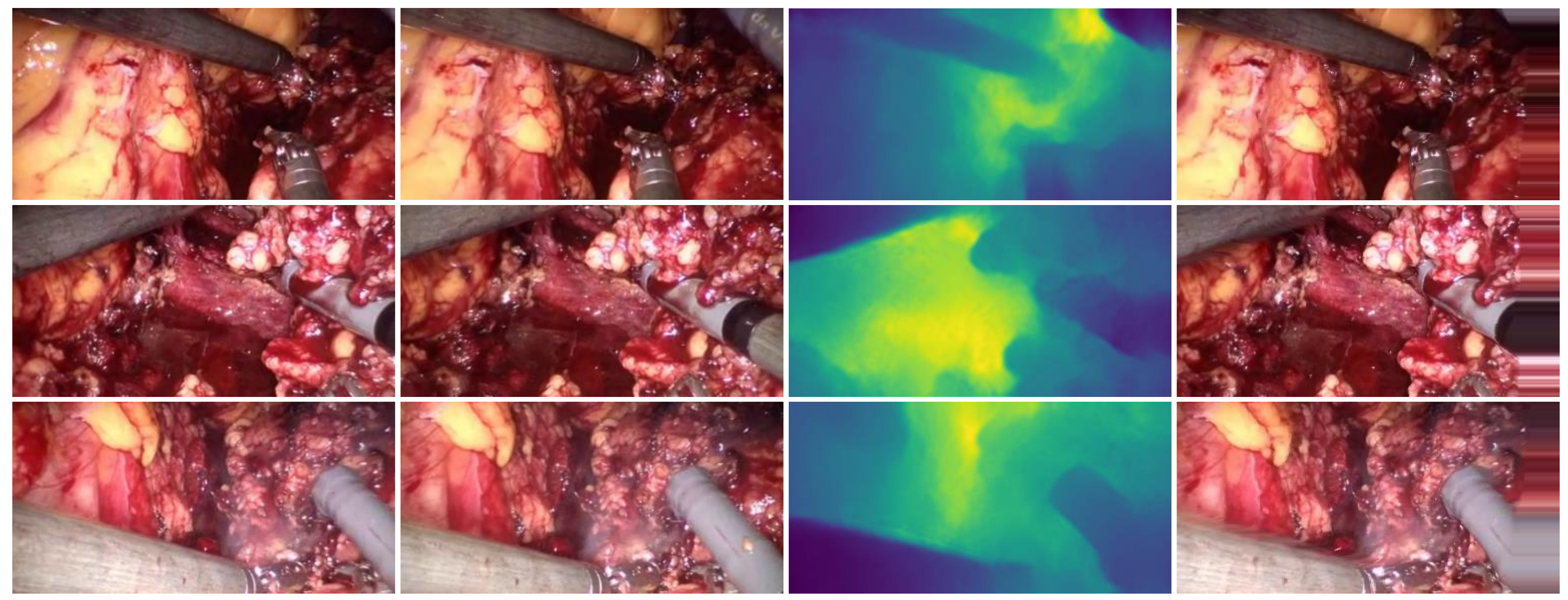}
\caption{Qualitative results on \textit{dVPN} dataset. From left to right, they are left image, right image, right depth map and reconstructed right image.} 
\label{network}
\end{figure*}

\begin{table*}[b]
\centering
\caption{SSIM score for \textit{dVPN} test set (higher is better).}\label{davinci_result}
\begin{tabular}{|p{3cm}<{\centering}|p{2.3cm}<{\centering}|p{1.8cm}<{\centering}|p{1.8cm}<{\centering}|}
\hline
Method & Training &Mean SSIM &Std. SSIM \\
\hline
ELAS \cite{geiger2010efficient}  &No training & 47.3  &0.079\\
SPS \cite{yamaguchi2014efficient}  &No training & 54.7 &0.092\\
V-Basic \cite{ye2017self} &Unsupervised  & 55.5 &0.106\\
V-Siamese \cite{ye2017self} &Unsupervised  & 60.4  &0.066\\
Monodepth \cite{godard2017unsupervised} &Unsupervised  &54.9   &0.087  \\
Monodepth2 \cite{godard2019digging}   &Unsupervised &71.2  & 0.075\\
SADepth (ours) &Unsupervised &\textbf{79.6} &\textbf{0.049}  \\ 
\hline
\end{tabular}
\end{table*}

\section{Experiments and results}
\subsection{Dataset}
We evaluated SADepth on two datasets. The first was the \textit{dVPN} dataset, collected from da Vinci partial nephrectomy, with 34320 pairs of rectified stereo images for training and 14382 pairs for testing \cite{ye2017self}. The second was the \textit{SCARED} dataset \cite{allan2021stereo} released during the Endovis challenge at MICCAI 2019, with 17206 pairs (dataset 1, 2, 3, 6 and 7) of rectified stereo images for training and 5637 pairs for testing. To verify the generalization of our framework, we only trained on the \textit{dVPN} dataset but test on both \textit{dVPN} and \textit{SCARED} dataset.

\subsection{Evaluation Metrics, Baseline, and Implementation Details}
\subsubsection{Evaluation Metrics} As the ground truth depth labels were not available for the \textit{in vivo} surgical data in the \textit{dVPN} dataset, we adopted the SSIM index to evaluate the similarity between the reconstructed image and the original input image (\textit{i.e.} \(\textbf{\textit{I}}^{r*}\) and \(\textbf{\textit{I}}^{r}\)) as the evaluation metric. For the \textit{SCARED} dataset the team at Intuitive Surgical collected the ground truth by using structured light, thus we used the absolute error to assess our SADepth model.

\subsubsection{Baseline}
We compared SADepth with several recent works. For the \textit{dVPN} dataset, we compared our method with stereo matching-based methods: ELAS \cite{geiger2010efficient} and SPS \cite{yamaguchi2014efficient}; Siamese-based networks: V-Basic \cite{ye2017self} and V-Siamese \cite{ye2017self}; and recent deep learning methods: Monodepth \cite{godard2017unsupervised} and the stereo mode of Monodepth2 \cite{godard2019digging}. For the \textit{SCARED} dataset, we compared our results with the methods summarized by the recent MICCAI sub-challenge paper \cite{allan2021stereo}.

\subsubsection{Implementation Details}
The SADepth model was implemented in PyTorch \cite{paszke2017automatic}, with a batch size of 16 and input/output resolution of \(192\times384\). The learning rate was set to \(10^{-4}\) for the first 15 epochs and then dropped to \(10^{-5}\) for the remainder. The model was trained for 20 epochs using the Adam optimizer which took about 22 hours on a single NVIDIA 2080 Ti GPU.

\begin{table*}[]
\centering
\caption{The mean absolute depth error for the SCARED test set 1 and 2 (unit: mm) (lower is better).}\label{Endovis_test}
\begin{tabular}{|p{4.0cm}<{\centering}|p{2.3cm}<{\centering}|p{1.7cm}<{\centering}|p{1.7cm}<{\centering}|}
\hline
Method  & Training  & \makecell[c]{Test Set 1 \\ Average} & \makecell[c]{Test Set 2 \\ Average}\\ \hline
Lalith Sharan \cite{allan2021stereo} &Supervised  &43.03 &48.72\\
Xiaohong Li \cite{allan2021stereo} &Supervised  &22.77 &20.52\\
Huoling Luo \cite{allan2021stereo} &Supervised  &19.52 &18.21\\
Zhu Zhanshi \cite{allan2021stereo} &Supervised  &9.60  &21.20\\
Wenyao Xia \cite{allan2021stereo} &Supervised  &6.73 &9.44\\

Congcong Wang \cite{allan2021stereo} &Supervised  &4.10 &4.28\\
Trevor Zeffiro \cite{allan2021stereo} &Supervised  &3.60 &3.47\\
J.C. Rosenthal \cite{allan2021stereo} &Supervised  &3.44  &4.05\\

Dimitris Psychogyios 1 \cite{allan2021stereo} &Supervised  &3.00  &1.67  \\
Dimitris Psychogyios 2 \cite{allan2021stereo} &Supervised &2.95  &2.30  \\
\hline
KeXue Fu \cite{allan2021stereo} &Unsupervised &20.94 &17.22  \\
Monodepth \cite{godard2017unsupervised} &Unsupervised  &23.56 &21.62  \\
Monodepth2 \cite{godard2019digging} &Unsupervised &21.92  &15.25  \\
SADepth (ours)  &Unsupervised   &\textbf{17.42}  &\textbf{11.23}  \\
\hline
\end{tabular}
\end{table*}

\subsection{Results}
The SADepth and other state-of-the-art results for the \textit{dVPN} dataset are summarized in Table~\ref{davinci_result} using the mean and standard deviation (Std.) of the SSIM index. The SADepth model effectively outperformed other methods with an SSIM of 79.6, \textit{i.e.} 24.7 units higher than Monodepth \cite{godard2017unsupervised},  8.4 units higher than Monodepth2 \cite{godard2019digging}, and 19.2 units higher than the Siamese architecture \cite{ye2017self}.

Table~\ref{Endovis_test} presents the results of SADepth on the test set 1 and test set 2 (as defined in the \textit{SCARED} dataset), together with the performance reported in the MICCAI sub-challenge summary paper \cite{allan2021stereo}. The results show an improvement over the unsupervised methods from the summary paper and recent baselines, while it is also competitive with some supervised approaches. This confirms that SADepth generalizes well across different datasets collected from different laparoscopes and subjects, while still producing superior performance compared with the state-of-the-art unsupervised approaches.

\section{Conclusions}
We have presented a new self-supervised adversarial depth estimation framework SADepth with an encoder-decoder generator and a concatenated stereo image pair as the input. The adversarial learning strategy improved the generation quality of the framework and led to the state-of-the-art performance on two public datasets. Furthermore, SADepth did not require any per-pixel depth labels and generalized well across different laparoscopes, suggesting excellent applicability to scalable data acquisition when accurate ground truth depth cannot be collected.

\bibliographystyle{splncs04}
\bibliography{reference}

\end{document}